\documentclass[preprint, amsmath]{revtex4-1}
\usepackage{latexsym}
\usepackage{amssymb}
\usepackage{amsfonts}

\usepackage{color}

\usepackage{graphicx}
\usepackage{amsmath}
\usepackage{subfigure}
\usepackage{times}
\usepackage{units}
\usepackage{hyperref}
\hypersetup{
  colorlinks = true,
  citecolor = blue,
  urlcolor = blue
}
\newcommand{{\bfr}}{\mbox{\boldmath$r$\unboldmath}}
\newcommand{{\bfv}}{\mbox{\boldmath$v$\unboldmath}}
\newcommand{{\bff}}{\mbox{\boldmath$f$\unboldmath}}
\newcommand{{\bfF}}{\mbox{\boldmath$F$\unboldmath}}
\newcommand{{\bfA}}{\mbox{\boldmath$A$\unboldmath}}
\newcommand{{\bfchi}}{\mbox{\boldmath$\chi$\unboldmath}}

\newcommand{{\cF}}{\mbox{\boldmath${\cal F}$\unboldmath}}
\newcommand{{\cG}}{\mbox{\boldmath${\cal G}$\unboldmath}}
\newcommand{{\cE}}{\mbox{\boldmath${\cal E}$\unboldmath}}
\newcommand{{\cB}}{\mbox{\boldmath${\cal B}$\unboldmath}}
\newcommand{{\cX}}{\mbox{\boldmath${\cal X}$\unboldmath}}
\newcommand{{\cY}}{\mbox{\boldmath${\cal Y}$\unboldmath}}
\def\v#1{{\bf#1}}

\begin{document}
\title{The wave equation in the birth of spacetime symmetries}
\author{Ricardo Heras}
\email{ricardo.heras.13@ucl.ac.uk}
\affiliation{Department of Physics and Astronomy, University College London, London WC1E 6BT, UK.}

\begin{abstract}
%\noindent
\noindent In 1887 Voigt published a paper dedicated to the Doppler effect in which he demanded form invariance to the wave equation in inertial frames and obtained a set of spacetime transformations now known as the Voigt transformations.  In 1905 Poincar\'e showed that the wave equation was also invariant under the Lorentz transformations. Voigt and Lorentz transformations are then closely related, but this relation is not widely known in the standard literature. In this paper we derive the Lorentz transformations from the invariance of the D'Alembert operator $\big(\Box^{2}=\Box'^{2}\big)$ and the Voigt transformations from the conformal invariance of the D'Alembert operator $\big(\Box^{2}=(1/\gamma^2)\Box'^{2},$ where $\gamma=1/\sqrt{1-v^2/c^2} \big).$   The homogeneous scalar wave equation is then invariant under the Lorentz transformations and conformally invariant under the Voigt transformations. We suggest a presentation of special relativity in which the Voigt transformations are commented after discussing the Galilean transformations but before presenting the Lorentz transformations.
\end{abstract}

\maketitle
\section{Introduction}
 \noindent Woldemar Voigt published in 1887 the article \cite{1}: ``On Doppler's Principle" which has received little recognition by physicists \cite{2,3,4,5,6}. Apparently, he was the first ---or at least one of the firsts--- who demanded form invariance of a physical law
to obtain a set of transformation equations. More precisely, Voigt demanded form invariance of the homogeneous wave equation in inertial frames and obtained a set of spacetime transformations now known as the Voigt transformations:
 \begin{eqnarray}
 x'=x -vt,\quad t'=t-\frac{vx}{c^2},\quad y'=\frac{y}{\gamma},\quad z'= \frac{z}{\gamma},
\end{eqnarray}
where $\gamma=1/\sqrt{1-v^2/c^2}$ is the Lorentz factor. Here we are adopting the standard configuration defined by two inertial frames $S$ and $S'$ in relative motion with speed $v$ along their common $xx'$ direction and whose origins coincide at the instant $t\!=\!t'\!=\!0$.

The reader will have noticed that the Voigt transformations are similar to the well-known Lorentz transformations of special relativity:
\begin{equation}
 x'=\gamma(x -vt),\quad t'=\gamma\bigg(t-\!\frac{vx}{c^2}\bigg), \quad y'=y, \quad z'=z.
\end{equation}
If the right-hand of the Voigt transformations in Eq.~(1) is multiplied by the factor $\gamma$ then the Lorentz transformations in Eq.~(2) are obtained. In 1905 Poincar\'e \cite{7} showed that the homogeneous wave equation is invariant under the transformations in Eqs.~(2). Therefore the Voigt and Lorentz transformations are closely related and it would be worthwhile to discuss this relation from a conceptual point of view.

Three brief comments enlighten the conceptual and historical importance of Voigt's 1887 paper: (i) Voigt transformations were obtained by the requirement of keeping the same form of the wave equation under inertial frames, and this is one application of which would be known later as the first postulate of special relativity. (ii) Form invariance of the wave equation carried the invariance of the speed of light, which constitutes the second postulate of special relativity. Interestingly, Voigt, without explicitly mentioning it, applied in practice the postulates of special relativity to the wave equation eighteen years before Einstein explicitly and concisely enunciated these postulates \cite{8}. (iii) As a consequence of the invariance of the speed of light: $c'=c$, the well-established Newtonian notion of absolute time: $t'=t$ should be replaced by the non-absolute time: $t'=t-vx/c^2$. According to Ives \cite{9} this was the first suggestion that: ``...a `natural' clock would alter its rate on motion.''
In this same sense Simonyi \cite{10} claims that when demanding form invariance of the wave equation, Voigt was ``...opening the possibility for the first time in the history of physics to call into question the concept of the absolute time.'' Voigt's non-absolute time was independently introduced in 1895 by Lorentz \cite{11} who called it  ``the local time.'' Interestingly, in a paper devoted to the Doppler effect, Voigt was inadvertently applying the postultes of special relativity. As pointed out by Ernst and Hsu \cite{2}: ``He was very close to suggesting a conceptual framework for special relativity.''

The reader might be surprised by the fact that Voigt's transformations are not usually mentioned in standard textbooks \cite{12}. His surprise would be even
greater when he could have observed that these transformations imply the same velocity transformation law of the special relativity:
\begin{equation}
 {\rm v}'_{x}=\frac{{\rm v}_{x}-v}{1-v{\rm v}_{x}/c^2},
\end{equation}
which is fully consistent with Voigt's assumption about the invariance of the speed of light, that is, if ${\rm v}_{x}=c$ is inserted in Eq.~(3) then it yields ${\rm v}'_{x}=c$. In addition, a further application shows that the Doppler effect predicted by the Voigt transformations turns out to be identical to that predicted by special relativity \cite{4,5}. The reader might be even more puzzled by the fact that Voigt obtained a set of transformations different from those of Lorentz, despite the fact that he essentially applied the two postulates of special relativity to the homogeneous wave equation.

Unfortunately, the Voigt transformations do not form a group \cite{3,4} and this seems to be the main reason for which these transformations have been overlooked in textbooks. Another reason seems to be the absence of a simple and pedagogical approach to derive these transformations.
The original derivation presented by Voigt \cite{1} is certainly not pedagogical nor easy-to-follow. In contrast, there are many pedagogical and easy-to-follow derivations of the Lorentz transformations, some of which are given in Ref. \cite{13}.

In this paper we hope to call attention to the Voigt transformations and add clarity to the close relation between Lorentz and Voigt transformations (a) by deriving the Lorentz transformations from the invariance of the D'Alembert operator $(\Box^{2}=\!\Box'^{2})$ and the Voigt transformations from its
conformal invariance $(\Box^{2}=(1/\gamma^2)\Box'^{2})$, (b) by pointing out that the homogeneous scalar wave equation is invariant under the Lorentz transformations and conformally invariant under the Voigt transformations, (c) by writing the Voigt transformations in the four-dimensional spacetime and showing that these transformations do not form a group, and (d) by suggesting a presentation of relativity in which the Voigt transformations are introduced after discussing the Galilean transformations and before presenting the Lorentz transformations. We do not present here a full discussion on
the physical or unphysical predictions of Voigt's transformations. We only point out some of these predictions briefly to emphasize the pedagogical usefulness of introducing these transformations as well as their historical and conceptual importance.

In Sec.~II we derive the Lorentz transformations from the invariance $\Box^{2}=\!\Box'^{2}$. In Sec. III we obtain the Voigt transformations from
the conformal invariance $\Box^{2}=(1/\gamma^2)\Box'^{2}$. In Sec.~IV we discuss the ideas of invariance and conformal invariance in the context of the
Lorentz and Voigt transformations. In Sec.~V we point out the relation between Lorentz and Voigt transformation in the four-dimensional spacetime and show that the latter do not form a group. In Sec.~VI we suggest the pedagogical presentation:
Galilean Trans. $\to$ Voigt Trans. $\to$ Lorentz Trans. In Sec.~VIII we emphasize the
 conceptual importance of the Voigt's 1887 paper. In the Appendix A we derive the Voigt transformations for the space and time derivative operators.

\section {The D'Alembert operator and the Lorentz transformations}

It is well-known that the D'Alembert operator is invariant under the Lorentz transformations. This result naturally suggests that the Lorentz transformations may be obtained from the invariance of the D'Alembert operator. Although this suggestion is correct, its explicit proof does not seem to be widely known in the standard literature. In this section we present such a proof which is inspired in one given by Parker and Schmieg (see Parker and Schmieg in Ref. \cite{13}).

In order to introduce notation and for future reference, we will first show that the D'Alembert operator is invariant under the Lorentz transformations.
For simplicity, we consider the standard configuration in which two inertial frames $S$ and $S'$ are in relative motion with speed $v$ along their common $xx'$ direction. The origins of the two frames coincide at the instant $t\!=\!t'\!=\!0$. The coordinates transverse to the relative motion of the frames $S$ and $S'$ are assumed to be invariant: $y'=y$ and $z'=z$. The corresponding derivative operators are also assumed to be invariant: $\partial /\partial y=\partial/\partial y'$ and $\partial /\partial z=\partial /\partial z'$. The D'Alembert operator is defined by $\Box^{2}\equiv\nabla^2\!-\!(1/c^2)\partial^2/\partial t^2$ in the frame $S$
and $\Box'^{2}\equiv\nabla'^2-(1/c^2)\partial^2/\partial t'^2$ in the frame $S'$. From the Lorentz transformations  $x'=\gamma(x -vt), $ and $t'=\gamma (t-vx/c^2)$, we derive the transformation laws for the derivative operators \cite{14}: $\partial/\partial x\!= \gamma(\partial/\partial x'-(v/c^2)\partial/\partial t')$ and $ \partial/\partial t= \gamma(\partial/\partial t'-\!v\partial/\partial x')$. It follows that
\begin{align}
\frac{\partial^2}{\partial x^2}=\gamma^2\bigg(\frac{\partial^2}{\partial x'^2}-\frac{2v}{c^2}\frac{\partial}{\partial x'}\frac{\partial}{\partial t'} +\frac{v^2}{c^4}\frac{\partial^2}{\partial t'^2}\bigg),\quad
\frac{\partial^2}{\partial t^2}=\gamma^2\bigg(\frac{\partial^2}{\partial t'^2}-2v\frac{\partial}{\partial t'}\frac{\partial}{\partial x'} +v^2\frac{\partial^2}{\partial x'^2}\bigg).
\end{align}
Using these transformations and $\partial^2 /\partial y^2=\partial^2/\partial y'^2$ and $\partial^2 /\partial z^2=\partial^2 /\partial z'^2$ we obtain
\begin{align}
\frac{\partial^2}{\partial x^2}+ \frac{\partial^2}{\partial y^2}  + \frac{\partial^2}{\partial z^2} -\frac{1}{c^2}\frac{\partial^2}{\partial t^2}=\gamma^2\bigg(1-\frac{v^2}{c^2}\bigg)\frac{\partial^2}{\partial x'^2}+ \frac{\partial^2}{\partial y'^2}  +\frac{\partial^2}{\partial z'^2}
-\frac{1}{c^2}\gamma^2\bigg(1-\frac{v^2}{c^2}\bigg)\frac{\partial^2}{\partial t'^2}.
\end{align}
From the definition of the factor $\gamma$ we have $\gamma^2(1-v^2/c^2)\equiv 1$ and therefore Eq.~(5) clearly shows the invariance of the D'Alembert operator: $\Box^{2}=\Box'^{2}$.

We will now travel the inverse route and demand the invariance of the D'Alembertian to obtain the Lorentz transformations. We consider again the standard configuration and assume $\partial /\partial y=\partial/\partial y'$ and $\partial /\partial z=\partial /\partial z'$. The invariance of the D'Alembertian can be expressed as
\begin{equation}
\frac{\partial^2}{\partial x^2}-\frac{1}{c^2}\frac{\partial^2}{\partial t^2}+\frac{\partial^2}{\partial y^2}  + \frac{\partial^2}{\partial z^2}=\frac{\partial^2}{\partial x'^2}-\frac{1}{c^2}\frac{\partial^2}{\partial t'^2}+ \frac{\partial^2}{\partial y'^2}  +\! \frac{\partial^2}{\partial z'^2}.
\end{equation}
A simple mathematical manipulation shows that Eq.~(6) can be expressed as
\begin{align}
\bigg(\frac{\partial}{\partial x}\!-\!\frac{1}{c}\frac{\partial}{\partial t}\bigg)\bigg(\frac{\partial}{\partial x}\!+\!\frac{1}{c}\frac{\partial}{\partial t}\bigg)\!+\!\frac{\partial^2}{\partial y^2} \!+\! \frac{\partial^2}{\partial z^2}=\bigg(\frac{\partial}{\partial x'}\!-\!\frac{1}{c}\frac{\partial}{\partial t'}\bigg)\bigg(\!\frac{\partial}{\partial x'}+\frac{1}{c}\frac{\partial}{\partial t'}\bigg)+ \frac{\partial^2}{\partial y'^2} \! +\! \frac{\partial^2}{\partial z'^2}.
\end{align}
By assuming linearity for the involved transformations of operators, we can write the quantities
\begin{align}
\bigg(\frac{\partial}{\partial x}-\frac{1}{c}\frac{\partial}{\partial t}\bigg)=A \bigg(\frac{\partial}{\partial x'}-\frac{1}{c}\frac{\partial}{\partial t'}\bigg),\quad
\bigg(\frac{\partial}{\partial x}+\frac{1}{c}\frac{\partial}{\partial t}\bigg)=A^{-1} \bigg(\frac{\partial}{\partial x'}+\frac{1}{c}\frac{\partial}{\partial t'}\bigg),
\end{align}
and $\partial /\partial y=\partial/\partial y'$ and $\partial /\partial z=\partial /\partial z'$. Insertion of these quantities in the left-hand side of  Eq.~(7) leads to an identity. The factor $A$ is independent of the derivative operators but can depend on the velocity $v$ and $A^{-1}=1/A$. In order to determine $A$, we demand that the expected linear transformation relating primed and unprimed time-derivative operators should appropriately reduce to the corresponding Galilean transformation \cite{15}: $\partial/\partial t= \partial/\partial t'-v\partial/\partial x'.$ Our demand is consistent with a linear transformation of the general form: $\partial/\partial t= F(v)(\partial/\partial t'\!-\!v\partial/\partial x'),$ where $F(v)$ depends on the velocity $v$ so that $F(v)\!\to\! 1$ when $v<<c$. From this general transformation it follows that if
$\partial/\partial t=0$ then $\partial/\partial t'=v\partial/\partial x'$ because $F(v)\neq 0$. Using this result in Eq.~(8) we obtain
\begin{align}
\frac{\partial}{\partial x}=A \bigg(\frac{\partial}{\partial x'}-\frac{v}{c}\frac{\partial}{\partial x'}\bigg),\quad
\frac{\partial}{\partial x}=A^{-1}\bigg(\frac{\partial}{\partial x'}+\frac{v}{c}\frac{\partial}{\partial x'}\bigg).
\end{align}
By combining these equations we can derive the expressions
\begin{equation}
A=\frac {\sqrt{1+v/c}  }{\sqrt{1-v/c} },\quad A^{-1}=\frac {\sqrt{1-v/c} }{\sqrt{1+v/c} },
\end{equation}
which can conveniently be written as
\begin{equation}
A=\gamma\bigg(1+\frac{v}{c}\bigg),\quad A^{-1}=\gamma\bigg(1-\frac{v}{c}\bigg).
\end{equation}
Insertion of these quantities into Eq.~(8) gives
\begin{align}
\bigg(\frac{\partial}{\partial x}\!-\!\frac{1}{c}\frac{\partial}{\partial t}\bigg)=\gamma\bigg(1\!+\!\frac{v}{c}\bigg) \bigg(\frac{\partial}{\partial x'}\!-\!\frac{1}{c}\frac{\partial}{\partial t'}\bigg),\,\,
\bigg(\frac{\partial}{\partial x}\!+\!\frac{1}{c}\frac{\partial}{\partial t}\bigg)=\gamma\bigg(1\!-\!\frac{v}{c}\bigg) \bigg(\frac{\partial}{\partial x'}\!+\!\frac{1}{c}\frac{\partial}{\partial t'}\bigg).
\end{align}
By adding and subtracting these equations, we obtain the transformation laws connecting unprimed and primed derivative operators
\begin{align}
\frac{\partial}{\partial x}= \gamma\bigg(\frac{\partial}{\partial x'}-\frac{v}{c^2}\frac{\partial}{\partial t'}\bigg),\quad
\frac{\partial}{\partial t}= \gamma\bigg(\frac{\partial}{\partial t'}-v\frac{\partial}{\partial x'}\bigg),
\end{align}
which must be completed with $\partial /\partial y=\partial/\partial y'$ and $\partial /\partial z=\partial /\partial z'$. These laws are the Lorentz transformations for derivative operators of the standard configuration. We note that the relations in Eq.~(13) imply coordinate transformations of the form: $x'=x'(x,t),$ $y'=y,$ $z'=z$ and $t'=t'(x,t).$ To find the explicit form of these transformations we can use Eq.~(13) to obtain
\begin{equation}
\frac{\partial x'}{\partial x}= \gamma,\quad
\frac{\partial x'}{\partial t}= -\gamma v,\quad \frac{\partial t'}{\partial t}= \gamma,\quad
\frac{\partial t'}{\partial x}=-\frac{\gamma v}{c^2}.
\end{equation}
The first relation in Eq.~(14)  implies (I): $x'=\gamma x + f_1(t),$ where $f_1(t)$ can be determined (up to a constant) by deriving (I) with respect to the time $t$ and using the second relation in Eq.~(14): $\partial x'/\partial t=\partial f_1(t)/\partial t=-\gamma v.$ This last equality implies (II):
$f_1(t)=-\gamma vt+ x_0,$ where $x_0$ is a constant. From (I) and (II) we obtain
\begin{equation}
x'=\gamma( x -vt) + x_0.
\end{equation}
The third relation in Eq.~(14) implies (III): $t'=\gamma t + f_2(x),$ where $f_2(x)$ can be obtained (up to a constant) from deriving (III) with respect to $x$ and using the last relation in Eq.~(14):
$\partial t'/\partial x=\partial f_2(x)/\partial x=-\gamma v/c^2$. This last equality implies (IV):
$f_2(x)= -\gamma v x/c^2 +t_0,$
where $t_0$ is a constant. From (III) and (IV) we obtain
\begin{equation}
t'=\gamma \bigg(t-\frac{vx}{c^2}\bigg) + t_0.
\end{equation}
The origins of the frames $S$ and $S'$ coincide at the time  $t\!=\!t'\!=\!0$ and therefore we have $x_0\!=\!0$ and $t_0\!=\!0$. In this way we obtain the Lorentz transformations of the standard configuration:
\begin{eqnarray}
 x'=\gamma(x -vt),\quad t'=\gamma \bigg(t-\frac{vx}{c^2}\bigg),
\end{eqnarray}
which must be completed with the remaining transformations: $y'=y$ and $z'=z.$
\section{The D'Alembert operator and the Voigt transformations}
 The Voigt conformal invariance of the D'Alembert operator is defined by the set of transformations that satisfy the relation \cite{16}
\begin{equation}
\Box^{2}=\frac{1}{\gamma^{2}}\Box'^{2}.
\end{equation}
We adopt again the standard configuration of the inertial reference frames $S$ and $S'$ but now we do not assume that the coordinates transverse to the relative motion of these frames are invariant. As a first step, we will verify that Voigt transformations in Eq.~(1) satisfy Eq.~(18).
In the Appendix A we explicitly show that the transformations in Eq.~(1) imply the following transformation laws \cite{4}: $\partial/\partial x=(\partial/\partial x'- (v/c^2)\partial/\partial t'), \partial/\partial t= (\partial/\partial t'-v\partial/\partial x'), \partial/\partial y=(1/\gamma)\partial/\partial x'$ and $ \partial/\partial z=(1/\gamma)\partial/\partial z'.$ From these relations it follows that
\begin{align}
\frac{\partial^2}{\partial x^2}=\frac{\partial^2}{\partial x'^2}-\frac{2v}{c^2}\frac{\partial}{\partial x'}\frac{\partial}{\partial t'} +\frac{v^2}{c^4}\frac{\partial^2}{\partial t'^2},\quad
\frac{\partial^2}{\partial t^2}=\frac{\partial^2}{\partial t'^2}-2v\frac{\partial}{\partial t'}\frac{\partial}{\partial x'} +v^2\frac{\partial^2}{\partial x'^2},
\end{align}
\begin{align}
\frac{\partial^2}{\partial y^2}=\frac{1}{\gamma^2}\frac{\partial^2}{\partial y'^2},\quad \frac{\partial^2}{\partial z^2} =\frac{1}{\gamma^2}\frac{\partial^2}{\partial z'^2}.
\end{align}
Making use of Eqs.~(19) and (20) we obtain
\begin{align}
\frac{\partial^2}{\partial x^2}+ \frac{\partial^2}{\partial y^2}  + \frac{\partial^2}{\partial z^2} -\frac{1}{c^2}\frac{\partial^2}{\partial t^2}=\frac{1}{\gamma^2}\!\bigg(\frac{\partial^2}{\partial x'^2}+\frac{\partial^2}{\partial y'^2}\!  +\! \frac{\partial^2}{\partial z'^2}
-\frac{1}{c^2}\frac{\partial^2}{\partial t'^2}\bigg).
\end{align}
This equation is the explicit form of Eq.~(18).

Now we will demand the covariance of the D'Alembert operator to derive the Voigt transformations. This derivation is similar to that proposed in the previous section to obtain the Lorentz transformations.  We consider again the standard configuration and assume Eq.~(18), or equivalently, Eq.~(21). A simple manipulation shows that Eq.~(21) can be expressed as
\begin{align}
\bigg(\!\frac{\partial}{\partial x}\!-\!\frac{1}{c}\frac{\partial}{\partial t}\!\bigg)\bigg(\frac{\partial}{\partial x}\!+\!\frac{1}{c}\frac{\partial}{\partial t}\!\bigg)\!+\!\frac{\partial^2}{\partial y^2}\!+\!\frac{\partial^2}{\partial z^2}\!=\!\frac{1}{\gamma^2}\Bigg(\!\bigg(\!\frac{\partial}{\partial x'}\!-\!\frac{1}{c}\frac{\partial}{\partial t'}\bigg)\bigg(\!\frac{\partial}{\partial x'}\!+\!\frac{1}{c}\frac{\partial}{\partial t'}\!\bigg)\!+\! \frac{\partial^2}{\partial y'^2}\!+\! \frac{\partial^2}{\partial z'^2}\!\Bigg).
\end{align}
By assuming linearity for the corresponding transformations of derivative operators, we can write
\begin{align}
\bigg(\frac{\partial}{\partial x}\!-\!\frac{1}{c}\frac{\partial}{\partial t}\bigg)=\frac{A}{\gamma}\bigg(\frac{\partial}{\partial x'}-\frac{1}{c}\frac{\partial}{\partial t'}\bigg),\quad
\bigg(\frac{\partial}{\partial x}\!+\!\frac{1}{c}\frac{\partial}{\partial t}\bigg)=\frac{A^{-1}}{\gamma} \bigg(\frac{\partial}{\partial x'}+\frac{1}{c}\frac{\partial}{\partial t'}\bigg),\quad
\end{align}
\begin{align*}
\frac{\partial}{\partial y}=\frac{1}{\gamma}\frac{\partial}{\partial y'},\quad
\frac{\partial}{\partial z}=\frac{1}{\gamma}\frac{\partial}{\partial z'},
\end{align*}
where the quantity $A$ is independent of the derivative operators but can depend on the velocity $v$. Following essentially the same argument leading to Eq.~(9) we can arrive at the expressions
 \begin{align}
\frac{\partial}{\partial x}=\frac{A}{\gamma}\bigg(\frac{\partial}{\partial x'}\!-\!\frac{v}{c}\frac{\partial}{\partial x'}\bigg),\quad
\frac{\partial}{\partial x}=\frac{A^{-1}}{\gamma}\bigg(\frac{\partial}{\partial x'}\!+\!\frac{v}{c}\frac{\partial}{\partial x'}\bigg).
\end{align}
Dividing these relations, we obtain the same equations for $A$ and $A^{-1}$ given in Eq.~(11). Therefore,  using Eq.~(11) in the first two relations displayed in Eq.~(23) we obtain
\begin{align}
\bigg(\frac{\partial}{\partial x}-\frac{1}{c}\frac{\partial}{\partial t}\bigg)=\bigg(1+\frac{v}{c}\bigg) \bigg(\frac{\partial}{\partial x'}-\frac{1}{c}\frac{\partial}{\partial t'}\bigg),\,\, \bigg(\frac{\partial}{\partial x}+\frac{1}{c}\frac{\partial}{\partial t}\bigg)=\bigg(1-\frac{v}{c}\bigg) \bigg(\frac{\partial}{\partial x'}+\frac{1}{c}\frac{\partial}{\partial t'}\bigg).
\end{align}
By adding and subtracting these equations, we obtain the corresponding transformation laws connecting unprimed and primed derivative operators, which are added to the transformation laws for $\partial/\partial y$ and $\partial/\partial z$ to obtain \cite{4}
\begin{align}
\frac{\partial}{\partial x}=\frac{\partial}{\partial x'}-\frac{v}{c^2}\frac{\partial}{\partial t'},\quad
\frac{\partial}{\partial t}= \frac{\partial}{\partial t'}-v\frac{\partial}{\partial x'},\quad \frac{\partial}{\partial y}=\frac{1}{\gamma}\frac{\partial}{\partial y'},\quad \frac{\partial}{\partial z}=\frac{1}{\gamma}\frac{\partial}{\partial z'}.
\end{align}
These are the Voigt transformations for derivative operators of the standard configuration. A direct manipulation of Eq.~(26) yields the corresponding inverse transformations \cite{4}:
\begin{align}
\frac{\partial}{\partial x'}=\gamma^2\bigg(\frac{\partial}{\partial x}+\frac{v}{c^2}\frac{\partial}{\partial t}\bigg),\quad
\frac{\partial}{\partial t'}= \gamma^2\bigg(\frac{\partial}{\partial t}+v\frac{\partial}{\partial x}\bigg),\quad \frac{\partial}{\partial y'}=\gamma\frac{\partial}{\partial y},\quad
\frac{\partial}{\partial z'}=\gamma\frac{\partial}{\partial z}.
\end{align}
The first two relations in Eq.~(26) imply coordinate transformations of the form: $x'=x'(x,t)$ and $t'=t'(x,t).$ To find the explicit form of these transformations we can use Eqs.~(26) to obtain
\begin{equation}
\frac{\partial x'}{\partial x}= 1,\quad
\frac{\partial x'}{\partial t}= -v,\quad \frac{\partial t'}{\partial t}= 1,\quad
\frac{\partial t'}{\partial x}=-\frac{v}{c^2}.
\end{equation}
From the first relation in Eq.~(28) it follows the equation (A): $x'= x + g_1(t),$ where $g_1(t)$ can be determined (up to a constant) by deriving (A) with respect to the time $t$ and using the second relation in Eq.~(28): $\partial x'/\partial t=\partial g_1(t)/\partial t=-v.$ This last equality implies (B):
$g_1(t)=-vt+ x_0,$ where $x_0$ is a constant. From (A) and (B) we obtain
\begin{equation}
x'=x -vt + x_0.
\end{equation}
The third relation in Eq.~(28) implies (C): $t'\!=\!t\! +\! g_2(x),$ where $g_2(x)$ can be obtained (up to a constant) from deriving (C) with respect to $x$ and using the last relation in Eq.~(28):
$\partial t'/\partial x\!=\!\partial g_2(x)/\partial x\!=\!-v/c^2$. This last equality implies (D):
$g_2(x)\!=\! -v x/c^2\!+\! t_0,$
where $t_0$ is a constant. From (C) and (D) we conclude
\begin{equation}
t'=t-\frac{vx}{c^2} + t_0.
\end{equation}
 The origins of the frames $S$ and $S'$ coincide at  $t\!=\!t'\!=\!0$. It follows that $x_0\!=\!0$ and $t_0\!=\!0$. In this way we obtain the Voigt transformations for the $x$ and $t$ coordinates of the standard configuration:
\begin{equation}
 x'=x -vt,\quad t'=t-\frac{vx}{c^2}.
\end{equation}
The transformations for the $y$ and $z$ coordinates are easily derived. From the last two relations in Eq.~(26) we obtain $\partial y'/\partial y=1/\gamma$ and $\partial z'/\partial z=1/\gamma$ which in turn imply  $y'=y/\gamma + y_0$ and $z'=y/\gamma + z_0$, where $y_0$ and $z_0$ are constants, which are vanished because the origins of the frames $S$ and $S'$ coincide at the time  $t=t'=0$. Thus
\begin{equation}
 y'= \frac{y}{\gamma} , \quad z'=\frac{z}{\gamma}.
\end{equation}
Equations~(31) and (32) are the Voigt transformations of the standard configuration. A direct manipulation of Eqs.~(31) and (32) yields to the corresponding inverse transformations \cite{5}:
\begin{align}
x=\gamma^2(x'+vt'),\quad t=\gamma^2\bigg(t'+\frac{vx'}{c^2}\bigg),\quad y=\gamma y',\quad  z=\gamma z'.
\end{align}
We note that the inverse transformations in Eqs.~(33) are obtained from the transformations in Eqs.~(31) and (32) by changing the roles of the primed and unprimed variables, reversing the sign of the velocity $v$ and multiplying the right-hand side of Eq.~(31) and (32) by the factor $\gamma^2$.

\section {Invariance vs Conformal invariance}
Lorentz and Voigt transformations are useful to illustrate the subtle difference between invariance and conformal invariance.
The next examples emphasize this difference. Using Eq.~(2) we can directly show the invariance
 \begin{equation}
 x'^2+y'^2+z'^2-c^2t'^2= x^2+y^2+z^2-c^2t^2.
\end{equation}
Evidently, the quantity: $x^2+y^2+z^2-c^2t^2$ preserves its form under the Lorentz transformations. Consider now the wavefront equation: $x^2+y^2+z^2-c^2t^2=0$, which describes a spherical light pulse sent out from the origin of the frame $S$ at $t=0$. From Eq.~(34) it follows that $x'^2+y'^2+z'^2-c^2t'^2=0$, which is the wavefront equation associated with a spherical light pulse sent out from the origin of the frame $S'$ at $t'=0$.
We conclude that the quantity  $x^2+y^2+z^2-c^2t^2$ is invariant under the Lorentz transformations and therefore the light wavefront equation $x^2+y^2+z^2-c^2t^2=0$ is also invariant under the same transformations.

Using Eq.~(1) we can directly show the result
\begin{equation}
 x'^2+y'^2+z'^2-c^2t'^2=\frac{1}{\gamma^2} \big(x^2+y^2+z^2-c^2t^2\big).
\end{equation}
The quantity $x^2+y^2+z^2-c^2t^2$ does not preserve its form under the Voigt transformations because of the presence of the conformal factor $1/\gamma^2$. However, Eq.~(35) states that the quantity $x'^2+y'^2+z'^2-c^2t'^2$ in the frame $S'$ is linear and homogeneously connected with the quantity $x^2+y^2+z^2-c^2t^2$ in the frame $S$. Quantities satisfying this kind of connections are generally called ``covariant'' and equations involving covariant quantities are called ``covariant equations.'' Voigt's conformal invariance is then a kind of covariance. Therefore we can say that the quantity $x^2+y^2+z^2-c^2t^2$ is ``conformally invariant'' or simply ``covariant'' under the Voigt transformations. If we consider again the light wavefront equation $x^2+y^2+z^2-c^2t^2=0$ in the frame $S$, then Eq.~(35) implies that $x'^2+y'^2+z'^2-c^2t'^2=0$ in the frame $S'$ because of $1/\gamma^2\not=0$. Thus the light wavefront equation is conformally invariant or covariant under the Voigt transformations.

We apply now the same order of ideas to the case of the D'Alembert operator. As already noted, this operator is invariant under the Lorentz transformations: $\Box^{2}=\!\Box'^{2}$. Consider now the
invariant scalar field $F(\v x,t)$, which satisfies $F(\v x,t)=F'(\v x',t')$. It follows that the quantity  $\Box^{2}F$ is Lorentz invariant: $\Box^{2}F=\Box'^{2}F'$. Therefore the homogeneous scalar wave equation is also Lorentz invariant. In short,
\begin{align}
{\rm If}\; \Box^{2}F=0\; {\rm then}\; \Box'^{2}F'=0\; {\rm because\; of} \,\,\Box^{2}= \Box'^{2}\; {\rm and}\;  F=F'.
\end{align}
We have pointed out that the D'Alembert operator is conformally invariant or covariant under the Voigt transformations: $\Box^{2}=(1/\gamma^2)\Box'^{2}$. If we consider again the field $F(\v x,t)$ then the quantity $\Box^{2}F$ is Voigt conformally invariant [or Voigt covariant]: $\Box^{2}F=(1/\gamma^2)\Box'^{2}F'$. The homogeneous scalar wave equation is then Voigt conformally invariant [or Voigt covariant]. In few words:
\begin{align}
{\rm If}\; \Box^{2}F=0\; {\rm then}\;\Box'^{2}F'=0\; {\rm because\; of}\; \, \Box^{2}=(1/\gamma^2) \Box'^{2},\, F=F'\; \,{\rm and}\;\, 1/\gamma^2\not=0.
\end{align}
Notice that the following statement: ``if an equation holds in an inertial frame then it must hold in every inertial frame'' is supported by the notions of invariance and covariance (or more precisely conformal invariance in our case). This explains why one can obtain two different sets of transformations by demanding the validity of the wave equation in two inertial frames. Let us elaborate on this point. In order to satisfy the demand: $\Box^{2}F=0$ in $S$ and $\Box'^{2}F'=0$ in $S'$, we can simply assume  $F=F'$ and  $\Box^{2}= \Box'^{2}$ and the latter leads to the Lorentz transformations. However, we can equally assume  $F=F'$ and  $\Box^{2}=(1/\gamma^2) \Box'^{2}$ and the latter leads to the Voigt transformations. Invariance and covariance (or conformal invariance) are connected with the same idea: ``unchange in form'' but in different forms.  Roughly speaking, covariance implies invariance but the latter does not imply the former.

\section{The relation between Lorentz and Voigt transformations}
The close relation between Lorentz and Voigt transformations is easily established using four-dimensional spacetime coordinates. Greek indices $\alpha, \beta,  \ldots$ run from 0 to 3; Latin indices $i,j,\ldots$ run from 1 to 3. Points are labeled as $x^{\alpha}=(x^0, x^1,x^2, x^3)=(ct,x,y,z)$ in the frame $S$ and $x'^{\alpha}=(x'^0, x'^1,x'^2, x'^3 )=(ct',x',y',z')$ in the frame $S'$. Summation convention on repeated indices is adopted. As is well-known, the Lorentz transformations in Eq.~(2) can be written as
\begin{equation}
x'^\alpha=\Lambda_\beta^\alpha x^\beta,
\end{equation}
where
\begin{equation}
\Lambda^\alpha_\beta=
  \begin{pmatrix}
    \gamma & -v\gamma/c & 0 &\quad 0\;\\
    -v\gamma/c & \gamma & 0 &\quad 0\;\\
     0 & 0& 1&\quad 0\;\\
    0& 0 & 0 & \quad 1\;\\
  \end{pmatrix},
\end{equation}
is the Lorentz matrix. Analogously, the Voigt transformations in Eq.~(1) can  be written as
\begin{equation}
x'^\alpha={\rm V}_\beta^\alpha x^\beta,
\end{equation}
where
\begin{equation}
{\rm V}^\alpha_\beta=
  \begin{pmatrix}
    1 & -v/c & 0 & 0\\
    -v/c & 1 & 0 & 0\\
     0 & 0& 1/\gamma & 0\\
    0& 0 & 0 & 1/\gamma\\
  \end{pmatrix},
\end{equation}
is the Voigt matrix. The relation between the Lorentz and Voigt matrices is given by
\begin{align}
\Lambda^\alpha_\beta=\gamma{\rm V}_\beta^\alpha,
\end{align}
that is, the Lorentz matrix is proportional to the Voigt matrix. Using Eq.~(42) we can find directly properties of the Voigt transformations from those of the Lorentz transformations.

The Lorentz matrices satisfy the relation $\Lambda^\alpha_\theta\Lambda^\theta_\beta=\delta^\alpha_\beta,$ where $\delta^\alpha_\beta$ is the kronecker delta.
This means that $\Lambda^\theta_\beta$ is the inverse of $\Lambda^\alpha_\theta$, which can be denoted as $(\Lambda^{-1})^\theta_\beta$. From $\Lambda^\alpha_\theta\Lambda^\theta_\beta=\delta^\alpha_\beta$ and Eq.~(42) we find
\begin{align}
{\rm V}^\alpha_\theta (\gamma^2{\rm V}^\theta_\beta)=\delta^\alpha_\beta.
\end{align}
Therefore $\gamma^2{\rm V}^\theta_\beta$ can be interpreted as the inverse of ${\rm V}^\alpha_\theta$, which can be denoted as ${\rm (V^{-1}})^\theta_\beta$. In its explicit form, this inverse reads
\begin{equation}
({\rm V^{-1}})^\theta_\beta=
  \begin{pmatrix}
    \gamma^2 & -v\gamma^2/c & 0 &\quad 0\;\\
    -v\gamma^2/c & \gamma^2 & 0 &\quad 0\;\\
     0 & 0& \gamma&\quad 0\;\\
    0& 0 & 0 & \quad \gamma\;\\
  \end{pmatrix}.
\end{equation}
The Lorentz matrices are defined to be those matrices satisfying $\Lambda^\mu_\alpha\eta_{\mu\nu}\Lambda^\nu_\beta=\eta_{\alpha\beta}$, where
\begin{equation}
\eta_{\alpha\beta}=
  \begin{pmatrix}
    1 & 0 & 0 & 0\\
    0& -1 & 0 & 0\\
     0 & 0& -1 & 0\\
    0& 0 & 0 & -1\\
  \end{pmatrix}.
\end{equation}
It follows that the Voigt matrices are defined to be those matrices satisfying
\begin{align}
{\rm V}^\mu_\alpha\gamma^2\eta_{\mu\nu}{\rm V}^\nu_\beta=\eta_{\alpha\beta}.
\end{align}
Despite the close relation between Lorentz and Voigt matrices, the latter do not form a group \cite{3,4}. To see this we can write Eq.~(46) in the more compact form: ${\rm V^T}\gamma^2\eta{\rm V}=\eta$, where ${\rm V^T}$ is the transposed matrix of ${\rm V}$ \big(notice that ${\rm V^{-1}}=\gamma^2{\rm V}^T\big)$.
We consider the Voigt matrices ${\rm V_1}$ and ${\rm V_2}$ to investigate if their product ${\rm V_1}{\rm V_2}$ is also another Voigt matrix (closure property). We have ${\rm V_1^T}\gamma_1^2\eta{\rm V_1}=\eta$ and ${\rm V_2^T}\gamma_2^2\eta{\rm V_2}=\eta$. Let ${\rm V_3}={\rm V_1}{\rm V_2}.$ Therefore ${\rm V_3^T}\gamma_3^2\eta{\rm V_3}=\big({\rm V_1^T}{\rm V_2^T}\big)\gamma_3^2\eta\big({\rm V_1}{\rm V_2}\big).$ If $\gamma_3=\gamma_1\gamma_2$ then
\begin{align}
{\rm V_3^T}\gamma_3^2\eta{\rm V_3}={\rm V_1^T}\gamma_1^2\big({\rm V_2^T}\gamma_2^2\eta{\rm V_2}\big){\rm V_1}={\rm V_1^T}\gamma_1^2\eta{\rm V_1}=\eta.
\end{align}
At first glance, it appears to be that the Voigt matrices satisfy the closure property. But this is not so because the assumption $\gamma_3=\gamma_1\gamma_2$ is incorrect. It can be shown that \cite{17}: $\gamma_3=\gamma_1\gamma_2\big(1+v_1v_2/c^2\big).$ Therefore the Voigt matrices do not satisfy the closure property.
Consequently, the Voigt transformations do not form a group since two successive Voigt transformations do not yield another Voigt transformations. This makes them unattractive from a physical point of view because they break the equivalence of the inertial frames.
As pointed out by L\'evy-Leblond (See L\'evy-Leblond 1975 paper in Ref. \cite{13}): ``The physical equivalence of the inertial frames implies a group structure for the set of all inertial transformations.'' However, we should argue in favour of Voigt's transformations that some of their predictions, like the transformation law for velocities and the formula for the relativistic Doppler effect, are also predictions of the Lorentz transformations \cite{18}.

\section{Some pedagogical comments}
As above pointed out, the Voigt transformations are not usually mentioned in standard textbooks despite the fact that these transformations can be derived from considerations of invariance in the wave equation. However, Voigt's transformations preserving the Galilean transformation $x'=x -vt$ and rejecting the absolute time $t'\not=t$, could be used as a previous step before introducing the Lorentz transformations. This means that the conventional presentation of the Lorentz transformations starting with the Galilean transformations could be conceptually improved by including an intermediate step in which the Voigt transformations are discussed. In a symbolic form, we suggest the following order of presentation
\begin{equation}
{\rm Galilean \;Trans.\; \longrightarrow Voigt\; Trans. \longrightarrow Lorentz \;Trans.}
\end{equation}
After all, the usual justification to introduce the Lorentz transformations is the failure of the Galilean transformations in preserving the form of the wave equation \big(See, for example,  R. K. Wagness in Ref.~\cite{15}, pp. 496 and 500;  and A. Zangwill in Ref.\cite{17},  pp. 824 and 829\big), failure that do not exhibit the Voigt transformations. The instructor could provide physical arguments why the Voigt transformations must be abandoned in favor of the Lorentz transformations, such as the lack of group structure in the Voigt transformations.

\section{The legacy of Voigt's 1887 paper}
Some brief reflections could explain the little impact caused by Voigt's 1887 paper among their contemporaries: (I) The main purpose of Voigt's paper was not to propose the ambitious idea of a new theory of space and time but simply to study the propagation of oscillating disturbances through an elastic incompressible medium and deduce the formula for the observed Doppler effect. (II) The process of finding a set of transformations that leave invariant the wave equation was not stressed in Voigt's paper as a new and fundamental idea. The only four words used by Voigt to justify this invariance were \cite{1}: ``as it must be'' (or ``da ja sein muss'' in the original German version). (III) Voigt did not provide any physical interpretation of his non-absolute time: $t'=t-vx/c^2$ nor commented anything about his proposal of the invariant character of the speed of light in inertial frames.
Apparently, Voigt  did not realized the great conceptual importance of his results. In connection with Voigt's ideas presented in his 1887 paper,
Hsu has pointed \cite{20}: ``If the physicists of the time had been imaginative enough, they might have recognized the potential of these ideas to open up a whole new view of physics.''

We usually find the names of Einstein, Poicar\'e, Lorentz, and Larmor in the creation and development of the special theory relativity. Voigt is relegated to be a minor contributor, in the best of cases. But this tradition is not faithful to the history of physics, since Voigt seems to have been the first in applying the two postulates of special relativity to the wave equation. The idea of demanding that the wave equation should not change its form when observed by different inertial frames was an important conceptual contribution. This is the legacy of Voigt's 1887 paper.
\appendix
\section{Derivation of Voigt's transformations for derivative operators }
\noindent Consider a function $F(x',y',z',t')$ in the frame $S'$. The total differential of $F$ reads
\begin{equation}
dF=\frac{\partial F}{\partial x'}dx' + \frac{\partial F}{\partial y'}dy' + \frac{\partial F}{\partial z'}dz' + \frac{\partial F}{\partial t'}dt'.
\end{equation}
The coordinates $x',y',z'$ and $t'$ are all functions of the coordinates $x,y,z$ and $t$ of the frame $S$. The total differentials of $x'$ and $t'$ read
\begin{align}
dx'&=\frac{\partial x'}{\partial x}dx+\frac{\partial x'}{\partial y}dy+\frac{\partial x'}{\partial z}dz+\frac{\partial x'}{\partial t}dt,\\
dt'&=\frac{\partial t'}{\partial x}dx+\frac{\partial t'}{\partial y}dy+\frac{\partial t'}{\partial z}dz+\frac{\partial t'}{\partial t}dt.
\end{align}
From the Voigt transformations in Eq.~(1) we have
\begin{align}
&\frac{\partial x'}{\partial x}= 1,\;\frac{\partial x'}{\partial y}= 0,\; \frac{\partial x'}{\partial z}= 0,\;\frac{\partial x'}{\partial t}=-v,\;
\frac{\partial t'}{\partial x}= -\frac{v}{c^2},\,
&\frac{\partial t'}{\partial y}= 0,\; \frac{\partial t'}{\partial z}= 0,\;
\frac{\partial t'}{\partial t}=1.
\end{align}
From Eqs.~(A2)-(A4) we obtain
\begin{align}
dx'=dx-vdt,\quad dt'=-\frac{v}{c^2}dx+dt.
\end{align}
By a similar procedure we can derive
\begin{align}
dy'=  \frac{dy}{\gamma},\quad dz'= \frac{dz}{\gamma}.
\end{align}
Using Eqs.~(A5) and (A6) into Eq.~(A1) and rearranging terms we have
\begin{align}
dF=\bigg(\frac{\partial F}{\partial x'}-\!\frac{v}{c^2}\frac{\partial F}{\partial t'}\bigg)dx+\frac{1}{\gamma}\frac{\partial F}{\partial y'}dy+\frac{1}{\gamma}\frac{\partial F}{\partial z'}dz+\bigg( \frac{\partial F}{\partial t'}-\!v\frac{\partial F}{\partial x'}\bigg)dt.
\end{align}
If now we consider $F$ as a function of the coordinates $x,y ,z$ and $t$ of the frame $S$ then the total differential can be written as
\begin{equation}
dF=\frac{\partial F}{\partial x}dx+\frac{\partial F}{\partial y}dy+\frac{\partial F}{\partial z}dz+\frac{\partial F}{\partial t}dt.
\end{equation}
Comparing the coefficients of  $dx,dy,dz$ and $dt$ in Eqs.~(A7) and (A8) we conclude that
\begin{align}
\frac{\partial F}{\partial x}=\frac{\partial F}{\partial x'}-\!\frac{v}{c^2}\frac{\partial F}{\partial t'}, \quad \frac{\partial F}{\partial t}= \frac{\partial F}{\partial t'}-\!v\frac{\partial F}{\partial x'},\quad \frac{\partial F}{\partial y}=\frac{1}{\gamma}\frac{\partial F}{\partial y'},\quad
\frac{\partial F}{\partial z}=\frac{1}{\gamma}\frac{\partial F}{\partial z'}.
\end{align}
If we drop $F$ in this equation then we obtain the Voigt transformations for the derivative operators.
Notice that we have derived these transformations two times. In the first derivation we demanded conformal invariance of the D'Alembertian and obtained Eq.~(26). In the second one, we directly use Eq.~(1) to derive Eq.~(A9).


\begin{thebibliography}{99}


\bibitem{1}
W. Voigt, ``\"Uber das Doppler'sche Princip," Nachr. Ges. Wiss. G\"ottingen, \textbf{8}, 41-51 (1887). An English version of this paper can be found in
Ref.~2. An English translateion is freely available in Wikisource:
\vskip 1pt \href{http://en.wikisource.org/wiki/Translation:On_the_Principle_of_Doppler}{http://en.wikisource.org/wiki/Translation:On\_the\_Principle\_of\_Doppler}




\bibitem{2}
A. Ernst, J.-P Hsu, ``First Proposal of the Universal Speed of Light by Voigt in 1887," Chinese J. Phys. \textbf{39}, 211-230 (2001).



\bibitem{3}
A. G. Gluckman, ``Coordinate Transformations of W. Voigt and the Principle of Special Relativity," \href{https://doi.org/10.1119/1.1974485}{Am. J. Phys. \textbf{36}, 226-231 (1968)}.



\bibitem{4}
A. G. Gluckman, ``Voigt Kinematics and Electrodynamic Consequences," \href{https://doi.org/10.1007/BF00708805}{ Found. Phys. \textbf{6}, 305-316 (1976)}.



\bibitem{5}
C. J.  Masreliez, ``Special relativity and inertia in curved spacetime," Adv. Studies Theor. Phys.\textbf{ 2}, 795-815 (2008).

\bibitem{6}
J. P. Wesley, ``Michelson-Morley result, a Voigt-Doppler effect in absolute space-time," \href{https://doi.org/10.1007/BF00735382}{Found. Phys. \textbf{16}, 817-824 (1986)}.

\bibitem{7}
H. Poincar\'e, ``Sur la Dynamique de l'\'Electron," Compt. Rend. Acad. Sci. Paris \textbf{140}, 1504-1508 (1905) and \href{https://doi.org/10.1007/BF03013466}{Rend. Circ. Mat. Palermo \textbf{21}, pp. 129-175 (1906)}. Excerpts of this paper in English appear in {\it The Genesis of General Relativity Boston Studies in the Philosophy of Science Vol.~3: Gravitation in the Twilight of Classical Physics: The Promise of Mathematics}. Eds. J. Renn and M. Schemmel (Springer-Verlag,  Netherlands, 2007), pp. 253-251. An English translation is freely available in Wikisource:
\vskip 1pt
\href{https://en.wikisource.org/wiki/Translation:On_the_Dynamics_of_the_Electron_(July)}{https://en.wikisource.org/wiki/Translation:On\_the\_Dynamics\_of\_the\_Electron\_(July)}.

\bibitem{8}
A. Einstein, ``Zur Elektrodynamik bewegter K\"orper," \href{https://doi.org/10.1002/andp.19053221004}{Ann. Phys. (Leipzig) \textbf{322}, 891-921 (1905)}. The English translation: ``On the Electrodynamics of Moving Bodies" can be found in {\it The Principle of Relativity} (Methuen, 1923, reprinted by Dover Publications, New York, 1952), pp. 35-65. An English translation is freely available in Wikisource: 
\vskip 1pt
\href{https://en.wikisource.org/wiki/On_the_Electrodynamics_of_Moving_Bodies_(1920_edition)}{https://en.wikisource.org/wiki/On\_the\_Electrodynamics\_of\_Moving\_Bodies\_(1920\_edition)}.

\bibitem{9}
H. E. Ives, ``Historical Note on the Rate of a Moving Clock", J. Opt. Soc. Am. \textbf{37}, 810-813 (1947).



\bibitem{10}
K. Simonyi, {\it A Cultural History of Physics} (CRC, Taylor and Francis, Boca Raton, Fl, 2012), p 405.

\bibitem{11}
H. A. Lorentz, \emph{Versuch einer Theorie der electrischen und optischen Erscheinungen in bewegten K\"orpern,}( E. J. Brill, Leiden 1895). An English translation is freely available in Wikisource:
\vskip 1pt
\href{http://en.wikisource.org/wiki/Translation:Attempt_of_a_Theory_of_Electrical_and_Optical_Phenomena_in_Moving_Bodies}{\footnotesize{https://en.wikisource.org/wiki/Translation:Attempt\_of\_a\_Theory\_of\_Electrical\_and\_Optical\_Phenomena\_in\_Moving\_Bodies}}


\bibitem{12}
Many recent textbooks do not discuss Voigt tranformations. Some books that mention these tranformations are, for example, R. N.
Henriksen, {\it Practical Relativity: From First Principles to the Theory of Gravity} (John Wiley 2011) problem 2.3; F. W. Searsand Brehme and R. W. {\it Introduction to the Theory of Relativity} (Addison-Wesley, Reading, Mass. 1968 ) p. 13; A. P. French, {\it Special Relativity} (Norton, New York 1968) p. 270; H. R. Brown, \emph{Physical Relativity: Space-time Structure from a Dynamical
Perspective} (Oxford University Press, New York 2005 ) p. 54.




\bibitem{13}
See, for example, L. Parker and G. M. Schmieg, ``Special Relativity and Diagonal Transformations,'' \href{https://doi.org/10.1119/1.1976289}{ Am. J. Phys. {\bf 38}, 218-222 (1970) };
J. M. Levy-Leblond, ``One more derivation of the Lorentz Transformation,'' \href{https://doi.org/10.1119/1.10490}{Am. J. Phys. {\bf 44}, 271-277 (1975)}; A. Macdonald, ``Derivation of the Lorentz transformation,'' \href{https://doi.org/10.1119/1.12487}{Am. J. Phys. {\bf 49}, 493 (1981) }; and J. M. L\'evy-Leblond, ``A simple derivation of the Lorentz transformation and of the accompanying velocity and acceleration changes,'' \href{https://doi.org/10.1119/1.2719700}{Am. J. Phys. {\bf 75}, 615-618 (2007) }.

\bibitem{14}
See, for example, W. G. V. Rosser, {\it Interpretation of Classical Electromagnetism} (Kluwer, Dordrecht, 1997), p. 378.

\bibitem{15}
See, for example, R. K. Wagness, {\it Electromagnetic fields} 2nd ed. (Wiley, New York, 1986), p. 496.


\bibitem{16}
The conformal group is generally defined as the group of transformations that leave the metric
invariant upto a scale: $g'_{\mu\nu}(x')= \lambda(x)g_{\mu\nu}(x)$, where we have used the conventional four-dimensional notation.
For the case of the Voigt transformations we have: $\eta'_{\mu\nu}= (1/\gamma^2)\eta_{\mu\nu}(x)$.

\bibitem{17}
See, for example, A. Zangwill, {\it Modern Electrodynamics} (Cambridge University Press, NY, 2013), p. 829.

\bibitem{18}
The authors of Ref.~2 have exhibited (but not discussed) a set of transformations which are, according to these authors, implied by the Voigt transformations and form a four-dimensional conformal group with two parameters.

\bibitem{19}
J. P. Hsu, {\it Einstein's Relativity and Beyond: New Symmetry Approaches.} Adavanced Series on Theoretical Physical Science, Vol. 7 (World Scientific, Signapore, 2000), p. 31.








\end{thebibliography}
\end{document}